\documentclass[12pt]{article}
    \usepackage{graphicx}
    \usepackage{dcolumn}
    \usepackage{bm}
    \usepackage[version=4]{mhchem}
    \usepackage{amsmath,amssymb}
    \usepackage{soul}
    \usepackage{siunitx}
    \usepackage{booktabs}
    \usepackage{graphicx}
    \usepackage{authblk}
    \usepackage{enumitem}
    \usepackage{lineno}
    \usepackage{caption}
    \usepackage{color}
    \usepackage{ulem}
    \usepackage{hyperref}

    \hypersetup{colorlinks=true, linkcolor=blue, citecolor=cyan, urlcolor=blue}
\title{Pressure-induced thermal expansion anomalies in dhcp iron hydride associated with magnetoelastic coupling}

\author[1,2]{Yuichiro Mori}
\author[1]{Katsutoshi Aoki}
\author[1]{Hiroyuki Kagi}
\author[1]{Masahiro Takano}
\author[3]{Ina Park}
\author[4]{Zifan Wang}
\author[4,5]{Duck Young Kim}
\author[6]{Noriyoshi Tsujino}
\author[6]{Sho Kakizawa}
\author[6]{Yuji Higo}  

\affil[1]{
    Graduate School of Science, The University of Tokyo, 
    7--3--1 Hongo, Bunkyo-ku, Tokyo 113--0033, Japan}
\affil[2]{
    The Department of Earth and Planetary Science, Faculty of Arts and Sciences, 
    Yale University, New Haven, Connecticut 06511, USA}
\affil[3]{
    Center for Computational Quantum Physics (CCQ), 
    Flatiron Institute, New York, NY 10010, USA}
\affil[4]{
    Center for High Pressure Science and Technology Advanced Research, 
    Shanghai 201203, People’s Republic of China}
\affil[5]{
    Division of Advanced Nuclear Engineering, Pohang University of Science and Technology,
    Pohang 37673, Republic of Korea}
\affil[6]{Japan Synchrotron Radiation Research Institute, SPring-8,
    1--1--1 Kouto, Sayo-cho, Sayo-gun, Hyogo 679--5198, Japan}

    \date{\today}

\begin{document}

\maketitle	
\thispagestyle{empty} 
\pagebreak
 \thispagestyle{empty}
 
\begin{abstract}
 Iron hydride with a double hexagonal close-packed structure (dhcp-FeH$_{x}$) undergoes a ferromagnetic-paramagnetic transition without changing its crystal structure.
Despite its relevance to metal-hydrogen interactions and magnetically driven elasticity, the extensive investigation of this phase is almost limited to room temperature.
Here, we performed XRD measurements at high pressure and high temperature, identifying the singularity in the temperature-volume relationship as the Curie temperature ($T_\text{C}$).
Pressurization lowered the $T_\text{C}$ of dhcp-FeH$_{x}$, and pronounced volume anomalies, indicating that pressure enhanced magnetoelastic coupling.
Density functional theory combined with dynamical mean-field theory (DFT+DMFT) reproduced the spontaneous magnetization and its negative pressure dependence of $T_\text{C}$, consistent with our experimental results. This establishes a methodology for determining magnetic transition temperatures and magnetoelastic coupling effects, and highlights dhcp-FeH$_{x}$ as a unique model system for providing new insights into itinerant-electron magnetism.
\end{abstract}
\pagebreak

\setcounter{page}{1}

\section{Introduction}
Hydrogen atoms occupy the interstitial sites of iron and form iron hydrides, 
expanding the interatomic distance and drastically altering the physical properties of metallic iron, 
including elasticity, magnetism, phase boundaries, and electronic properties via Fe-H interactions~\cite{fukai2006metal,ishimatsu2012hydrogen,meier2019pressure}
The physical properties of iron hydrides have attracted keen attention as a representative transition-metal hydride 
for understanding the interaction between 3\textit{d} transition metals and hydrogen atoms. 
The change in the physical property of iron by hydrogenation is also essential for the Earth and planetary sciences because hydrogen turns into a siderophile element at high-pressure and high-temperature (high-$PT$) conditions~\cite{okuchi1997hydrogen, tagawa2021experimental} 
and the solubility of H in Fe, which is the planetary core-forming element, becomes significantly large~\cite{fukai1984iron, badding1991high}.

The hydrogenation of transition metals alters their magnetic properties. Among these, iron, when hydrogenated from its high-pressure phase, retains its overall crystal structure but exhibits changes in its magnetic properties. 
While these changes in magnetism can strongly affect the volume of some materials, there has been limited research on the magnetostriction of hydrides. 
On the other hand, this issue is a primary concern when considering hydrogen-metal interactions. 
In general materials, while understanding magnetic contributions to thermal expansion has progressed~\cite{lohaus2023thermodynamic}, the effect of pressure on magnetoelastic properties remains poorly understood. 
To date, no study has experimentally resolved how magnetism couples to volume at high-$PT$, nor whether magnetic transitions produce measurable thermodynamic signatures. 
This long standing gap persists despite the central role of Fe-H interactions in condensed-matter physics and deep planetary interiors.

\begin{figure}[htbp]
    \centering
    \includegraphics[width=0.5\textwidth]{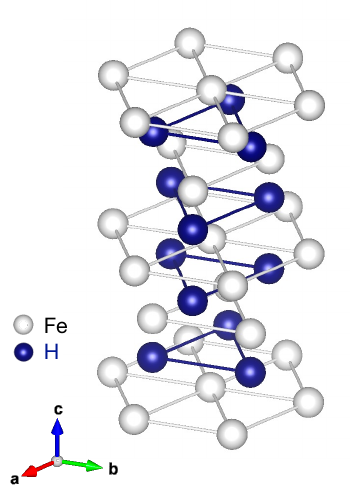}
    \caption{Cystal structure of dhcp-FeH$_x$. 
    Iron atoms occupy 2\textit{a} and 2\textit{c} sites, and Hydrogen atoms dissolve into 4\textit{f} sites in the Wyckoff representation. 
    The atomic radius ratio of hydrogen and iron is distorted, and the first nearest-neighbor atoms in each layer are connected by lines to clearly visualize the stacking sequence (ABAC stacking). 
    The space group of dhcp-FeH$_x$ is the same as the high-pressure phase (hcp) of Fe with a hexagonal close-packed structure, but the stacking order is different from the hcp structure (ABAB stacking). 
    This stacking alternation results in the unit cell of dhcp-FeH$_x$ being twice as large along the \textit{c}-axis as that of hcp.}
    \label{fig:xtal}
\end{figure}

Iron hydride with double hexagonal close-packed structure (dhcp-FeH, Fig.~\ref{fig:xtal}) is a particularly intriguing target 
for investigating metal–hydrogen interaction because it hosts pressure-dependent magnetism in a wide pressure and temperature ($PT$)-range~\cite{sakamaki2009melting,pepin2014new}. 
Previous experimental studies have shown that dhcp-FeH$_x$ remains ferromagnetic up to $\sim$30~GPa at ambient temperature, 
as determined by M\"{o}ssbauer spectroscopy and X-ray magnetic circular dichroism (XMCD) measurements~\cite{ishimatsu2012hydrogen, mao2004nuclear}. 
Density functional theory calculations predict stability up to 45–60~GPa at 0~K~\cite{elsasser1998ab2}. 
These results suggest that the Curie temperature ($T_\text{C}$) decreases with increasing pressure.

In addition to that, some calculations predicted an enormous magnetic contribution to volume, indicating its potential to show the thermal expansion anomaly such as the volume invariant to temperature or the negative thermal expansion (NTE). 
Because of the unique features of wide-$PT$ stability and the potentially large volume induced by magnetism, volume measurements of dhcp-FeH$_x$ at high-$PT$ conditions can be particularly advantageous for direct comparison with experimental results and theoretical calculations. 
Despite its ideal characteristics, the magnetic properties and their contribution to elastic properties in dhcp-FeH$_x$ have never been examined due to experimental difficulties. 
Existing constraints on ($T_\text{C}(P)$) remain purely computational~\cite{gomi2018effects} or are limited to low temperature experiments~\cite{ying2020magnetic}.

Here, we reveal a previously unidentified singularity in the temperature-volume relations of dhcp-FeH$_x$ at high pressures 
and demonstrate that this anomaly arises from magnetic volume collapse associated with the magnetic transition. 
DFT+DMFT calculations of spontaneous magnetization corroborate this interpretation, showing that the anomaly occurs at the corresponding $T_\text{C}$. 
These results uncover a magnetic transition in FeH, captured directly as anomalies in thermal expansion, 
providing the first experimental evidence of magnetically induced volume anomalies in a transition-metal hydride.

\section{Methods}
\subsection{Experimental Procedures}
\begin{figure}[htbp]
    \centering
    \includegraphics[width=\textwidth]{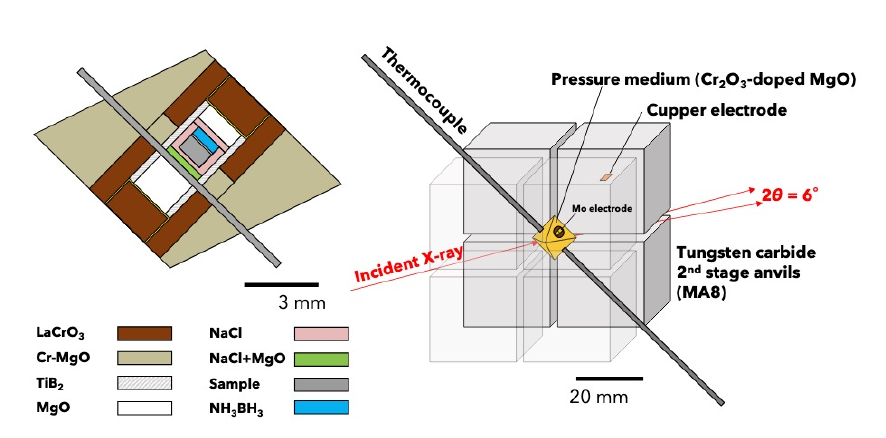}
    \caption{Cell assembly of XRD measurements under high-$PT$ conditions. 
    The sample and \ce{NH3BH3} were separated by hexagonal boron nitride (hBN) disk. 
    Molybdenum foil is used as electrode connecting the anvils to the cylindrical heater. 
    Octahedron pressure medium with 10~mm edge length was set into the MA8 (right-hand side of the figure) 
    and compressed by the six first-stage anvils attached to DIA-type guide-block. 
    The temperature was recorded every 10 seconds by a digital multi-meter of the electromotive force between a pair of the thermocouple inserted nearby the sample capsule.
}\label{figS:cell}
\end{figure}

In situ energy-dispersive X-ray diffraction experiments under high-$PT$ conditions were conducted at BL04B1 (SPring-8) and NE7A (PF-AR, KEK). 
Pressure was generated using a Kawai-type multi-anvil apparatus with DIA-type guide blocks installed on the uniaxial presses, ’SPEED-Mk.II’~\cite{katsura2004large} and ’MAX-III’ installed at BL04B1 and NE7A, PF-AR, KEK, respectively. 
In the cell assembly, the pelletized iron powder and ammonia borane (\ce{NH3BH3}) were encased together (Fig.~\ref{figS:cell}).

\begin{figure}[htbp]
    \centering
    \includegraphics[width=\textwidth]{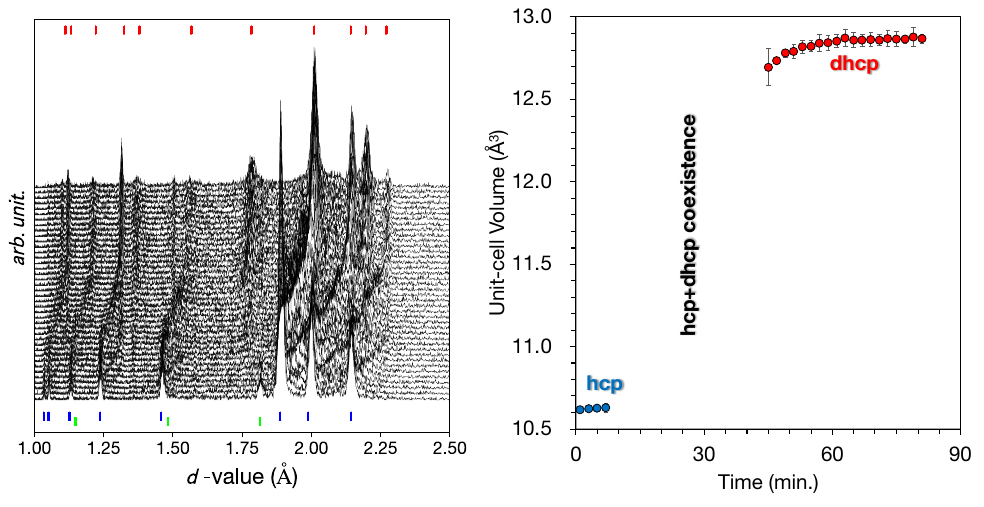}
    \caption{In-situ synthesis of FeH at 15~GPa and 850~K.
    The left figure shows the sequentially obtained raw X-ray diffraction profile during hydrogenation. 
    From bottom to top, hcp iron transformed into dhcp structure by hydrogenation. 
    The exposure time of each profile is 120 seconds. 
    It took around 80~minutes until the volume expansion induced by hydrogenation ceased. 
    Lower green and blue ticks represent the peaks from surrounding NaCl and hcp-FeH$_{x}$, respectively. 
    Upper red ticks represent the peaks from dhcp-FeH$_x$.
    The right figure shows the time-resolved unit-cell volume change of iron hydride. 
    In the initial stage, hcp-structured FeH$_{x}$ formed. 
    As hydrogenation proceeded, the hydrogen content exceeded the solubility of hcp-FeH$_{x}$. 
    Then, dhcp-FeH$_{x}$ appeared alongside saturated hcp-FeH$_{x}$. 
    In the final stage, hcp-FeH$_{x}$ vanished, and dhcp-FeH$_{x}$ stopped expanding as the hydrogen in iron reached saturation.}
    \label{fig:hyd}
\end{figure}

The sample and the hydrogen source were enclosed together in a NaCl capsule, which is known empirically to be an effective hydrogen-sealing material under high-$PT$ conditions. 
The detailed experimental procedures are as follows:
\begin{enumerate}[label = (\roman*)]
    \item The sample was compressed to 15~GPa, in which the hcp phase is stable for pure Fe~\cite{takahashi1964high,klotz2008alpha}.
    \item The temperature was increased to 850~K to decompose the hydrogen source (\ce{NH3BH3}) and induce the formation of dhcp-FeH$_x$. 
    Time-resolved X-ray diffraction (XRD) measurements were collected every two minutes to monitor the phase transition and volume expansion induced by hydrogenation (Fig.~\ref{fig:hyd}). 
    After the unit-cell volume reached a constant value, XRD profiles of the sample and pressure marker were collected at 850~K and the pressure was determined.
    \item The temperature was then gradually decreased from 850~K to 300~K over 180 minutes while acquiring time-resolved XRD data of dhcp-FeH$_x$. Each diffraction pattern was accumulated for 4~minutes. 
    The temperature of each pattern was determined as the average over its accumulation period (temperature was recorded every 10 seconds).
    The pressure for each XRD profile was assumed to decrease linearly with decreasing temperature. This assumption was validated by preliminary experiments.
    \item After completing step (iii), the sample was compressed to $\sim$18~GPa at 300~K and heated to 850~K. XRD profiles of the sample and pressure marker right after temperature stabilization.
    \item Steps (iii) and (iv) were repeated under different applied loads.	
\end{enumerate}
Using this procedure, sequential temperature-dependent XRD measurements were conducted at approximately 16, 18, 20, and 22~GPa. 
For phase identification and lattice parameter determination, the X-ray diffraction profiles were analyzed using PDIndexer software~\cite{seto2010development}. 

\subsection{Pressure determination}
$P$–$T$ paths depend on the material of the sample assemblies, the temperature range, and the heating itinerary.
With this assembly, the pressure decreases almost linearly from the pressure before and after each temperature-decreasing path evident from the preliminary experiment, 
which measured every 50~K with the same assembly and the same pressure regions.
We succeeded in obtaining ‘snapshot-like’ XRD by slowly cooling during time-resolve XRD measurements. 
Temperature change during each measurement is less than 15~K, and pressure change is less than 0.05 GPa by applying the $P$–$T$ linear approximation.

\subsection{The detail of cell assemblies}
In the cell assembly (Fig.~\ref{figS:cell}), 
Fe pellet mixed with hexagonal BN as a$_{x}$ grain-growth inhibitor~\cite{saitoh2017p} and a hydrogen source, ammonia borane (\ce{NH3BH3})~\cite{nylen2009thermal}, 
were loaded into the NaCl capsule. 
It is empirically known that 2/3 of the prepared H/Fe ratio is used to hydrogenate the sample. 
To make dhcp-FeH$_{x}$ with $x \sim 1$, the starting H/Fe ratio was set to 2, 
giving that the molar content of released hydrogen was predicted to be 1.3 times larger than that of iron where hydrogen content is in excess. 
Pressure and temperature were respectively determined using the pressure marker made of NaCl mixed with MgO grain-growth inhibitor just below the NaCl capsule 
and a pair of W3\%Re-W25\%Re thermocouples. 
Pressure was determined using the EoS of NaCl~\cite{matsui2012simultaneous}. 
A cylindrical \ce{TiB2} + BN + AlN composite heater were used~\cite{kanzaki2010crystal}. 
The sample heating is accomplished by the changing the input electrical power and the temperature was monitored by emf of the inserted thermocouple. 
By decreasing the applied electrical power, corresponding to 850~K, to 0~W for 180~min and the XRD profile of dhcp-FeH$_x$ were sequentially obtained during decreasing temperature.

\subsection{DFT+DMFT calculations for clarifying $PT$-dependence of magnetism}
In DFT+DMFT calculations, the crystal structure parameters, the unit-cell volume at each pressure condition and $c/a$ ratio, 
were adopted from the experimental reports~\cite{hirao2004compression} for the whole temperature range of calculation (Table~\ref{tab:dmft-structure}). 
A constant $c/a = 3.270$ was used for the pressure range investigated via DFT+DMFT.
The charge self-consistent DFT+DMFT calculation was performed as implemented in the DFT + embedded DMFT Functional (eDMFT) code~\cite{Haule2007, haule2010dynamical}. 
The DFT calculation with full-potential augmented plane wave method was performed using WIEN2k code~\cite{Blaha2018}, where the Perdew-Burke-Ernzerhof (PBE) generalised gradient approximation (GGA) was used for the exchange-correlation functional~\cite{Perdew1996}. 
A $19\times19\times5$ k-point mesh was used for the electronic self-consistent calculation. For the DFT+DMFT calculations, the hybridization energy window ranges from -10~eV to 10~eV at $P = 4$~GPa. 
The corresponding band indices range were fixed for higher pressure conditions. 
Slater parametrization was applied with a Coulomb interaction parameter $U = 5.7$~eV and Hund’s exchange parameter $J = 0.8$~eV for Fe 3\textit{d} orbitals~\cite{han2018phonon}. 
The impurity model was solved using a continuous-time quantum Monte Carlo (CTQMC) impurity solver~\cite{haule2010dynamical, Georges1996}.

\section{Results}
\subsection{Experimental constraints on Curie temperature and magnetostriction}
\begin{figure}[htbp]
    \centering
    \includegraphics[width=\textwidth]{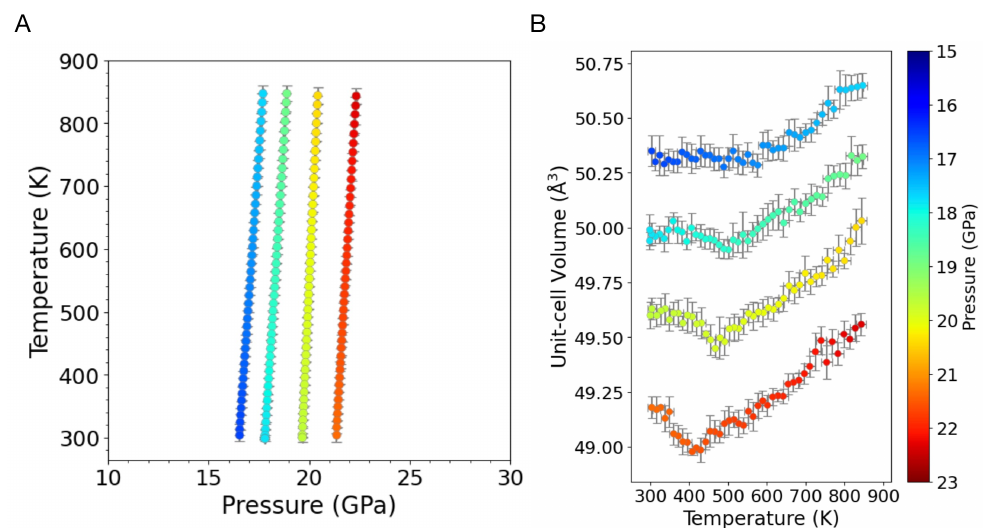}
    \caption{Pressure–Temperature–Volume ($P$–$T$–$V$) relations of dhcp-FeH$_x$.
    (A) $P$–$T$ conditions for volume measurements by the energy-dispersive method with white X-ray. 
    In this regime, only dhcp-FeH$_x$ was present. 
    (B) $T$–$V$ relation of dhcp-FeH, showing the thermal expansion anomaly, 
    the volume invariant to temperature at 15–16~GPa, and exhibiting the negative thermal expansion at 17–23~GPa. 
    All data points were obtained in the temperature-decrease pathway and contoured by pressure in both figures.}
    \label{fig:PTV}
\end{figure}

After synthesizing dhcp-FeH$_x$ at 15~GPa and 1000~K (Fig.~\ref{fig:hyd}), 
we conducted a sequential in situ XRD measurement of dhcp-FeH$_x$ during the cooling process at 16, 18, 20, and 22~GPa (Fig.~\ref{fig:PTV}A). 
The temperature-volume ($T$–$V$) relation exhibited clear discontinuities (Fig.~\ref{fig:PTV}B). 
Given that hydrogen content ($x$) is a continuous quantity of temperature ($T$)~\cite{fukai2006metal}, 
these discontinuities are unlikely to arise from changes in $x$. 
More precisely, the pressure fluctuates during the measurement. 
In the assembly used and within the experimental temperature range, 
the pressure difference between the highest and lowest temperatures is about 1–1.5~GPa (Fig.~\ref{fig:PTV}A), 
which cannot explain these anomalies. Those anomalies, therefore, suggest an underlying phase transition. 
However, the XRD profiles of dhcp-FeH$_x$ showed no change in crystal structure, 
indicating that this remarkable $T$–$V$ behavior is likely due to the interplay between magnetic volume and thermal expansion.

\begin{figure}[htbp]
    \centering
    \includegraphics[width=\textwidth]{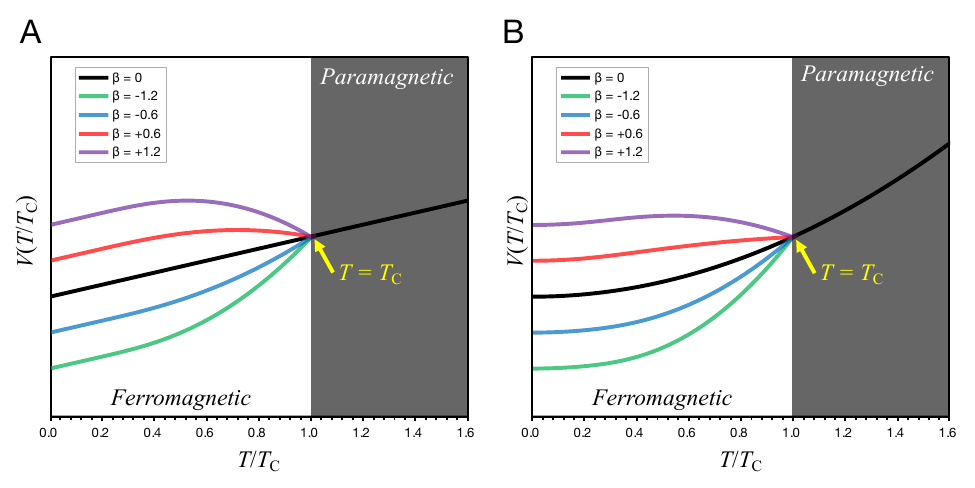}
    \caption{$T$–$V$ relations with strong magnetostriction in two simple models. 
    (A) Volume is assumed to be described as $V(T/T_\text{C}) = \alpha(T/T_\text{C}) + \beta \alpha M_\text{S}(T/T_\text{C})^{2}$, 
    where $\alpha$ and $\beta$ are constants and $M_\text{S}$ is spontaneous magnetization, which depends on $T/T_\text{C}$. 
    Magnetic contribution to volume is proportional to the squared $M_\text{S}$~\cite{wohlfarth1977thermodynamic, shiga1988magnetism}.
    (B) Volume is assumed to be described as $V(T/T_\text{C}) = \alpha(T/T_\text{C})^{2} + \beta \alpha M_\text{S}(T/T_\text{C})^{2}$.
    Material with a positive $\beta$ can exhibit Invar behavior and negative thermal expansion. 
    In either case, the rapid decrease in $M_\text{S}$ near $T_\text{C}$ causes the transition point to appear as a singular point in $T$–$V$.}\label{figS:TVbend}
\end{figure}

The singular point in the $T$–$V$ relation through the ferromagnetic-paramagnetic transition corresponds to the Curie temperature ($T=T_\text{C}$) 
because the magnetic contribution to the volumes of ferromagnetic materials is proportional to the square of the spontaneous magnetization~\cite{wohlfarth1977thermodynamic, shiga1988magnetism}. 
Figure~\ref{figS:TVbend} shows a simple model for describing the modeled $T$–$V$ relation that undergoes a ferromagnetic-paramagnetic transition across $T_\text{C}$. 

Although the emergence of a visible magnetic contribution to the volume (magnetostriction) depends on the strength of the magnetoelastic interaction, the phase transition point ($T_\text{C}$) always becomes a singular point. 
It will appear as a kink in the $T$–$V$ relation. Bending of the $T$–$V$ associated with the ferromagnetic-paramagnetic phase transition has been observed in bcc iron~\cite{ridley1968lattice} and in the Invar alloy at $T=T_\text{C}$~\cite{shiga1996invar}. 
To extract the magnetic contribution to the volume from the thermal expansion, it is necessary to determine $T_\text{C}$ at each pressure. 
Except for the lowest-pressure path, the thermal expansion shows negative values (negative thermal expansion, NTE). 
The sequential XRD profiles at $P\sim$21~GPa are shown in Fig.~\ref{figS:Index} to illustrate the NTE of dhcp-FeH$_x$. At the lowest pressure, the $T$–$V$ relation is nearly constant up to $T_\text{C}$, indicating Invar-like behavior 
(i.e., the magnetic contribution to the volume is opposite to the thermal expansion, and both effects are comparable). 
The $T$–$V$ discontinuity is exhibited just below 600~K, but it is difficult to precisely determine the singular point in the $T$–$V$ relation due to potential errors in volume measurements. 
Thus, we determined $T_\text{C}$ along three pressure paths --- 18, 20, and 22~GPa.

To estimate the magnetostriction, an equation of state in the paramagnetic region must be constructed and interpolated into the ferromagnetic region, 
since in the ferromagnetic regime, the observed temperature dependence of the unit-cell volume is a sum of thermal expansion and magnetism-induced volume change. 
We applied the second-order Birch-Murnaghan equations of state (EoS) with the Mie-Gr\"{u}neisen-Debye (MGD) approximation to fit the $PVT$ dataset above 600~K, 
which is well above the observed $T_\text{C}$. The resultant $V_0$, $K_0$ and $\gamma_{0}$ are 54.81(13)~\AA$^{3}$, 147(2) GPa, and 2.6(1), respectively by fixing Debye temperature to 420~K~\cite{cornell1997inelastic} and $q = 1$. 
Note that $V_0$ is the hypothetical unit-cell volume of ferromagnetic dhcp-FeH, and $K_{0}$ is the isothermal bulk modulus. 
$\gamma_{0}$ and $q$ are the Gr\"{u}neisen parameter and the dimensionless parameter used to construct the MGD-EoS, respectively. 
The subscript '0' of those elastic parameters refers to the standard condition, 
where $P = 0$~GPa and $T = 300$~K in this study. 
The difference between the calculated volume by this EoS and the observed volumes above 600~K is less than 0.5\%. 
Note that this EoS is constructed using only in the narrow pressure range, 
and extrapolation to substantially higher pressure should be treated with caution. 

\subsection{Pressure dependence of Curie temperature of dhcp-FeH$_x$}
\begin{figure}[htbp]
    \centering
    \includegraphics[width=\textwidth]{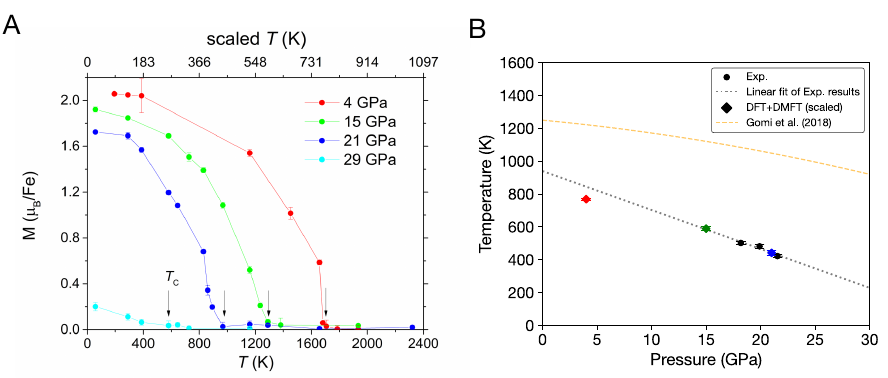}
    \caption{Magnetization and curie temperature estimated by DFT+DMFT calculations. 
    (A) DFT+DMFT results on the Temperature dependence of magnetization. 
    The bottom-X axis shows raw temperature values used in the DMFT calculation, and the top-X axis shows the scaled temperature, $rT$. 
    The arrow marks the Curie temperature at which the magnetization becomes zero. 
    Error bars are the standard deviation of the magnetization from the converged Monte Carlo calculation results. (B) $P$–$T_\text{C}$ of dhcp-FeH$_x$ obtained from DFT+DMFT calculations and experiments. 
    Note that $T_\text{C}$ at 29~GPa is excluded because the temperature dependence of the magnetization deviates significantly from the typical Ginzburg–Landau behavior 
    (The plots including data up to 29~GPa are shown in the Fig.~\ref{figS:critical}A). 
    The orange dashed line represents the previously reported $P$–$T_\text{C}$ derived by Korringa-Kohn-Rostoker method with coherent potential approximation (KKR-CPA) calculation~\cite{gomi2018effects}. 
    Black solid circles indicate the $T_\text{C}$ measured in experiment, and the dotted line represents the linear ferromagnetic-paramagnetic boundaries of each path determined 
    by fitting the high temperature part (paramagnetic part) of the $T$–$V$ data as a linear function of temperature and interpolated to lower temperature to obtain $T_\text{C}$. 
    The error bars on the vertical axis represent the temperature change during a single measurement.
    The $P$–$T_\text{C}$ relation obtained from calculations is normalised. }
    \label{fig:PTC_DMFT}
\end{figure}

We performed first-principles density functional theory combined with dynamical mean-field theory (DFT+DMFT) calculations 
to investigate the FM and PM states of dhcp-FeH$_x$ under various high-$PT$ conditions. 
Crystal structure of dhcp-Fe used in this simulation is listed in Table~\ref{tab:dmft-structure}. 
DFT+DMFT calculation enables not only approaching the finite temperature conditions but also treating the electronic correlation effect of Fe d electrons by describing dynamical quantum fluctuations, 
in which the latter is revealed to be significant for dual (itinerant-localised) nature of magnetic iron~\cite{vonsovskii1993localized,hausoel2017local}.
Electronic correlation effects are expected to be non-negligible at high temperature and/or pressure~\cite{lichtenstein2001finite, katanin2010orbital, leonov2011electronic, pourovskii2013electronic}.
We investigated temperature-magnetization relation at wider range of pressure conditions --- 4, 15, 21, and 29~GPa (Fig.~\ref{fig:PTC_DMFT}A).
We emphasize that the DFT+DMFT calculations are intended to assess the consistency with the experimentally observed pressure dependence of the magnetic transition, rather than to reproduce the absolute values of the $T_\text{C}$. It is noteworthy that DFT+DMFT usually overestimate the value of  $T_\text{C}$, and the
degree of overestimation depends on several factors including the type of Coulombic
interaction form. In this calculation, we used the so-called ’Ising-type’, which includes only the density-density type of interaction.
This type usually overestimates the $T_\text{C}$ by a factor of 2 to 3, and it also depends on the value of interaction parameters $U$ and Hund’s coupling $J_\text{H}$, 
while giving a reasonable value of magnetization and temperature dependence of magnetization for the overall temperature range~\cite{han2018phonon, lichtenstein2001finite}.
Therefore, while the pressure dependence of magnetization at the normalized temperature and relative change in $T_\text{C}$ with respect to pressure can be directly compared, 
the absolute value of the phase transition temperature requires scaling to check the consistency with the experimental values.

We first linearly fitted the $P$–$T_\text{C}$ of experimental values from 18–23~GPa and obtained the relation 
$T_\text{C}^\text{exp.} (\text{K}) = -23.7\cdot P(\text{GPa}) + 939.6$ 
because $T_\text{C}$ is postulated to decrease linearly with pressure ($R^{2} = 0.92$) over the experimental pressure range (18–23~GPa). 
After that, we introduced the temperature scaling factor, $r$, 
which is defined as the ratio between the $T_\text{C}^\text{DMFT}$  and $T_\text{C}^\text{exp.}$  at $P = 21$~GPa. 
In our calculation, the $T_\text{C}$ was overestimated by a factor of around 2.2 (i.e., the temperature axis was scaled by the factor of $r = 0.457$). 
Other pressure conditions were also tested for scaling, and the scaling factor, $r$ was between $0.44$–-$0.48$, 
indicating that the linear approximation of pressure and $T_\text{C}$ holds well within the experimental range. 
Linear extrapolation to ambient pressure yields $T_\text{C}$ of dhcp-FeH$_x$ to be 940~K, close to that of bcc-Fe, 1053~K. 
Figure~\ref{fig:PTC_DMFT}B and show the pressure dependence of $T_\text{C}$.
In contrast, DFT+DMFT calculations suggest that $P$–$T_\text{C}$ trend deviate from the linear relation at 4~GPa and significantly collapse at 29~GPa (Fig.~\ref{figS:29GPa}A). 
In particular, the calculated magnetic moment at 29~GPa remains finite but small with the value of $M_\text{S}=0.2$~µB at $T=58$~K, 
and the scaled $T_\text{C}$ was obtained as 258~K. 
At this pressure the temperature dependence of the magnetization deviates significantly from the typical Ginzburg–Landau behavior (Fig.~\ref{figS:29GPa}B).

\section{Discussion}

\subsection{Hydrogen content in dhcp-FeH$_x$}

\begin{figure}[htbp]
    \centering
    \includegraphics[width=0.6\textwidth]{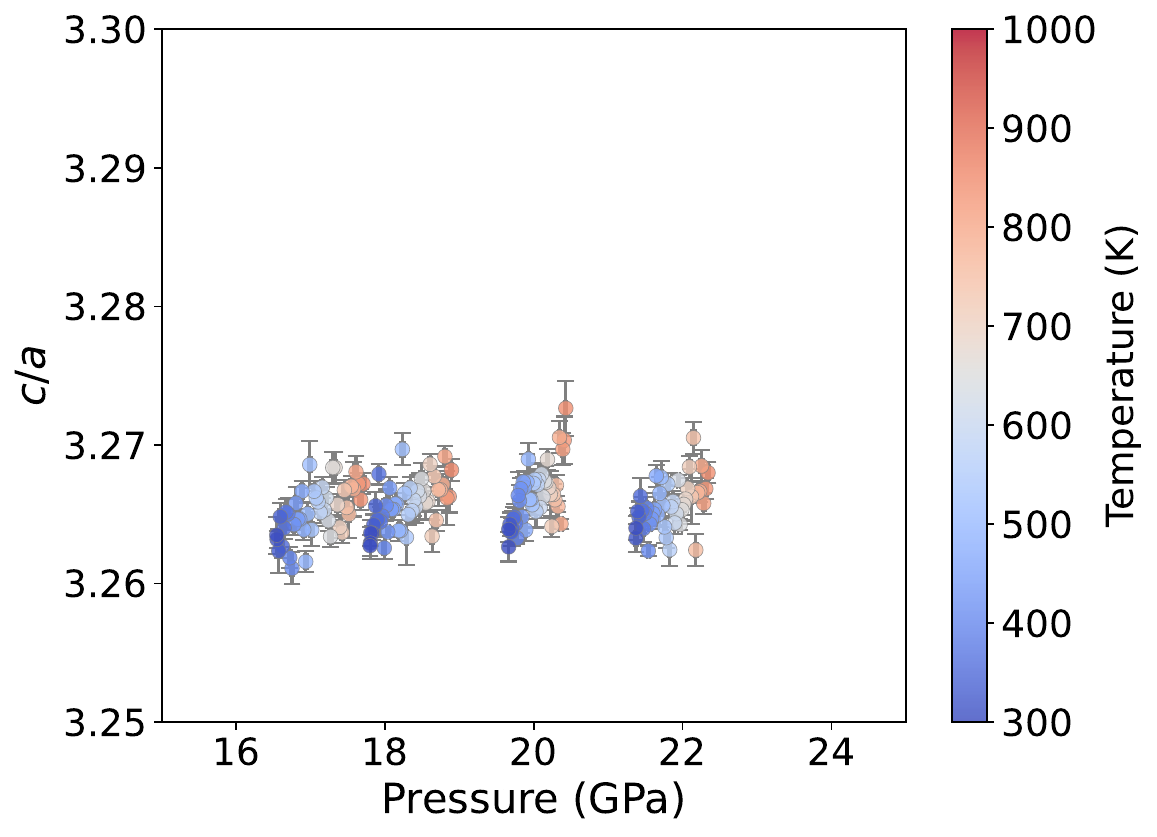}
    \caption{Pressure dependence of $c/a$ ratio of dhcp-FeH$_x$ countoured by temperature.
  }
    \label{fig:c_a}
\end{figure}

Because hydrogen content can modify the magnetic properties of dhcp-FeH$_x$ ~\cite{gomi2018effects, tsumuraya2012first}, it is essential to constrain the hydrogen content in dhcp-FeH$_x$. 
In iron hydrides, the hydrogen content incorporated into the interstitial sites has commonly been estimated by comparing the EoS of Fe and FeH$_x$ with the same crystal structures.
Although a high-$T$ EoS of dhcp-FeH$_x$ has not yet been established, 
its compression behavior at 300~K has been investigated by previously X-ray diffraction studies~\cite{pepin2014new,hirao2004compression}.
Because pure Fe does not exhibit the dhcp structure as a stable phase, we applied the EoS of hcp‑Fe as a counterpart (reference volume of hcp‑Fe) to estimate hydrogen content and volume change by hydrogenation.
Note that this choice is justified because dhcp‑FeH shares the same space group $P6_{3}/mmc$ with hcp‑Fe, and the stacking sequence of Fe planes has twice the periodicity of hcp‑Fe along the $c$ axis (Fig.~\ref{fig:xtal}).
The experimentally determined $c/a$ ratio is 3.2657(18), 
where the value in parentheses denotes the standard deviation of all measured data points. The uncertainty propagated from the measurement errors at each data point is approximately half of this value. In the examined $PT$-range, no significant $PT$- dependence of the $c/a$ ratio was observed (Fig.~\ref{fig:c_a}), and the value is similar to the previous reports of compression studies of FeH at 300~K~\cite{pepin2014new,hirao2004compression}.

Combined with atomic volumes of hcp-Fe~\cite{yamazaki2012p} and dhcp-FeH~\cite{pepin2014new} calculated from their compression curves, we estimated the hydrogen-indunced volume expansion. 
At 0~GPa, the atomic volume of dhcp‑FeH is approximately 1.26 times larger than that of hcp‑Fe, 
and this ratio decreases to about 1.19 at 50~GPa.
Within the pressure range in the present study (16–22~GPa), the volume ratio derived from the 300~K compression curves is 1.216(3).
In comparison, our experimentally obtained dhcp‑FeH dataset yields a volume ratio of 1.210(4).
These values agree within experimental uncertainty, indicating a consistent volume expansion of approximately 20\% relative to hcp‑Fe in the experimental range.

The volume expansion by one hydrogen atom per iron atom ($v_\text{H}$) is often used as a standard for estimating the hydrogen content in metal hydrides~\cite{fukai2006metal}. By expressing the volume expansion of dhcp-FeH with $v_\text{H}$, the derived value at 0~GPa is 2.16(5)~\AA$^3$/H-atom and it decreases to 1.90~\AA$^3$ at 50~GPa.
This value is in excellent agreement with those reported for iron hydrides with hcp and fcc structures, for which typical $v_\text{H}$ expansions are $\sim$2.0–2.5~\AA$^3$.
Applying the hydrogen‑induced volume change derived from the 300~K compression curve directly to our experimental volume data yields a hydrogen content of $x = 0.98(2)$.
In principle, $v_\text{H}$  may exhibit some temperature dependence. 
However, such temperature effects would be significantly suppressed at higher pressures. 
Therefore, it is reasonable to apply the $v_\text{H}$ at 300~K to the present high‑temperature dataset as a first‑order approximation. From these considerations, we conclude that the hydrogen content in dhcp-FeH$_x$ is close to stoichiometric (i.e., $x \sim 1$) throughout the present experiments and remains nearly constant  within experimental uncertainty. Importantly, the volume change associated with hydrogen incorporation is more than an order of magnitude larger than the volume change induced by magnetism discussed below.
This observation demonstrates that the observed $T$-$V$ discontinuities is not likely to be attributed to changes in hydrogen content. Accordingly, all subsequent analyses and discussions are carried out by assuming stoichiometric dhcp‑FeH ($x \sim 1$).

\subsection{Magnetostriction of dhcp-FeH$_x$}
\begin{figure}[htbp]
    \centering
    \includegraphics[width=\textwidth]{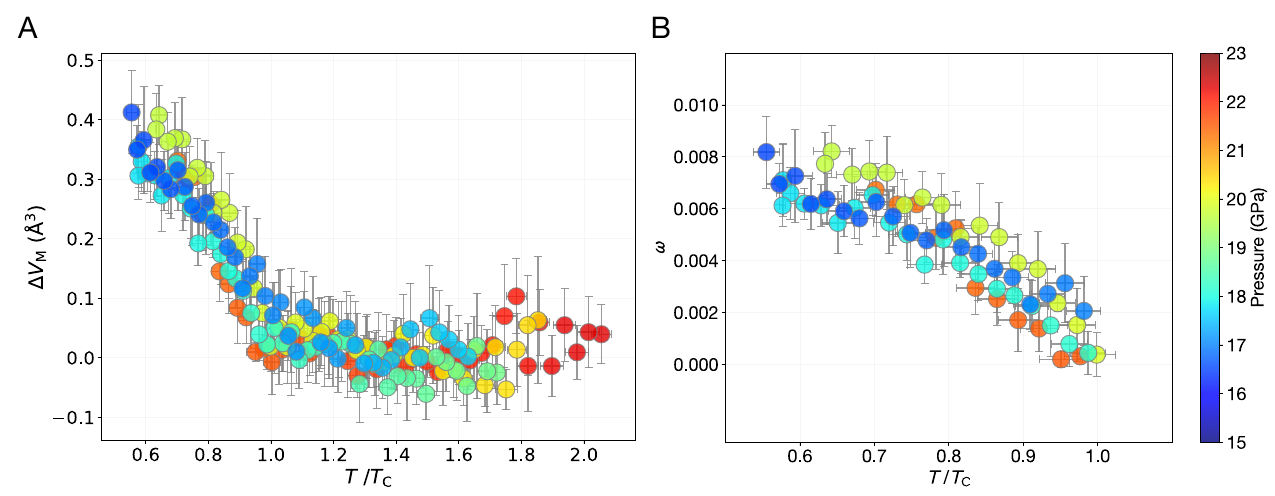}
    \caption{Temperature dependence on the magnetic effect on volume.
    (A) Temperature dependence on volume expansion induced by ferromagnetism ($\Delta V_\text{M}$). 
    $T_\text{C}$ was calculated using the formula,
    $T_\text{C}(\text{K}) = -23.7\cdot P(\text{GPa}) + 939.6$ . 
    The magnetic volume contribution ($\Delta V_\text{M}$) change was estimated by subtracting the calculated paramagnetic volume using EoS constructed above $T_\text{C}$ from the measured those values in the ferromagnetic region.
    Figure~\ref{figS:Vmag_diff} shows the graphical methodology to estimate the $\Delta V_\text{M}$ from the $T$–$V$ data.
    (B) The relation between temperature normalised by $T_\text{C}$ and spontaneous volume magnetostriction, $\omega$, which is equal to $\Delta V_\text{M}/V$. 
    The data is excluded for the part of $T>T_\text{C}$ to visualise the relation between pressure and spontaneous magnetostriction.
  }
    \label{fig:T_Vmag}
\end{figure}

Figure~\ref{fig:T_Vmag}A shows the relation between temperature normalised by $T_\text{C}$ 
and the magnetic contribution to the unit-cell volume ($\Delta V_\text{M}$).
The contribution to the unit-cell volume via magnetism ($\Delta V_\text{M}$) at each point corresponds to the difference between 
the observed volume and the hypothetical paramagnetic volume.
The paramagnetic unit-cell volume was calculated from the experimentally obtained EoS constructed in the paramagnetic regime, 
as discussed above. 
Figure~\ref{figS:Vmag_diff}A shows the $T$–$V$ calculated from paramagnetic dhcp-FeH$_x$ EoS with with observed dataset. 
The relationship between temperature and the difference in these values (i.e., $\Delta V_\text{M}$) is shown in Fig.~\ref{figS:Vmag_diff}B.

In studies of Invar alloys, spontaneous magnetostriction, $\omega$, is frequently used to evaluate the strength of the magnetic contribution. 
This value is defined as $\Delta V_\text{M}$ normalised by its unit-cell volume. 
Spontaneous magnetostriction ($\omega$) is described as the product of the magnetoelastic coupling constant, 
the compressibility, and the square of the spontaneous magnetization ($M_\text{S}$)~\cite{wohlfarth1977thermodynamic, shiga1988magnetism}. 
Therefore, combined with mean-field approximation, the magnetization of Fe atoms in dhcp-FeH$_x$ can be estimated as a function of temperature ($S = 1/2$). 
The magnetic properties of dhcp-FeH$_x$ at room temperature were examined experimentally by neutron diffraction at $P = 4.2$~GPa~\cite{saitoh2020crystal} 
and by XMCD intensity up to $\sim$30~GPa~\cite{ishimatsu2012hydrogen}. 
The spontaneous magnetization ($M_\text{S}$) just below  is described as 
$M_\text{S} \propto (T_\text{C}-T)^{\beta}$, 
where $\beta$ is a critical exponent with universality. 
Applying mean-field approximation, we can estimate the critical exponent, $\beta$, individually from $\Delta V_\text{M}$ (or $\omega$) and $T/T_\text{C}$. 
Figure~\ref{figS:critical} shows the estimated $M_\text{S}$ and the observed magnetorstirction at experimental points below $T_\text{C}$. 
The derived critical exponent $\beta$ estimated from magnetostriction ($\omega = V_\text{M}/V$) ranges from 0.39 to 0.47, 
and this value is 80–90\% of the value calculated from $T_\text{C}$ using the mean-field approximation. 
In classical theories, such as the Ginzburg–Landau theory and the mean (molecular) field approximation, $\beta$ is equal to 0.5 in adjacent to $T=T_\text{C}$. 
In contrast, typical experimental values of are around 0.33~\cite{kittel2018introduction}, which is rather close to the three-dimensional Ising model and Heisenberg model than 0.5, 
suggesting short-range correlations. Our results suggest that we can interpret the $\Delta V_\text{M}$–$T/T_\text{C}$ curves as reflecting a ferromagnetic–paramagnetic transition with small spin fluctuations. 
We successfully reproduce both the volume difference and the estimated $T_\text{C}$ by incorporating the magnetostriction expression into a mean-field approximation model.
Figure~\ref{fig:T_Vmag}B shows the $T/T_\text{C}$–$\omega$ in the ferromagnetic region ($T<T_\text{C}$). 
Those plots overlap within the experimental errors, indicating the magnetostriction of dhcp-FeH$_x$ is independent on pressure. 
In contrast, $T_\text{C}$, which corresponds to the singular point in $T$–$V$ plots, decreases with increasing pressure. 
Because magnetoelastic coupling effect ($C$) is denoted as $C = \omega / \kappa M_\text{S}^{2}$~\cite{wohlfarth1977thermodynamic, shiga1988magnetism}, 
where $\kappa$ is compressibility of material. 
The right-hand side numerator, $\omega$, we can use the obtained values in $T$–$V$ measurements. 
In the denominator, $\kappa$ is defined as inverse of bulk modulus ($K$), and $K$ at each ($P, T$) conditions can be determined by using paramagnetic dhcp-FeH$_x$ EoS under an assumption 
that $K(P, T)$ of ferromagnetic dhcp-FeH$_x$ is same as that of paramagnetic one. 
For magnetic properties, we first constructed the relation between pressure and $M_\text{S}(P)$ at room temperature by combining the pressure dependence of magnetization at up to 30~GPa~\cite{ishimatsu2012hydrogen} 
with and the magnetic structure analysis at 4.2~GPa~\cite{saitoh2020crystal}. 
Since direct observations for the magnetic property measurements of dhcp-FeH$_x$ have not been performed above room temperature, 
we applied the mean-field approximation to express $M_\text{S}$ as a function of $T/T_\text{C}$.

The magnetization ($M_\text{S}$) at $T/T_\text{C} \sim$ 0 divided by $M_\text{S}$ ($T/T_\text{C} \sim 0.5$) was about 1.2 at 15~GPa and 1.4 at 21~GPa, respectively.
As shown in Fig.~\ref{fig:T_Vmag}B, the calculated $\Delta V_\text{M}$ from the experimental results is 0.5~\AA$^3$ at $T/T_\text{C} \sim 0.5$. 
With using the mean-field approximation and the relation between magnetic moment and magnetostriction, $\Delta V_\text{M}$ at 0~K get 0.7~\AA$^3$ at 15~GPa and 1.0~\AA$^3$ at 21~GPa. 
Those values are significantly smaller than the previously calculated estimates of 4.0~\AA$^3$~\cite{fukai2006metal, elsasser1998ab2}. 
Here, we will compare this value to the hydrogen-induced volume expansion. 
In 3\textit{d} transition metals, the volume expansions induced by one H atom incorporation normalized by the number of host atoms fall into 2.0–2.5~\AA$^3$ per one hydrogen (deuterium) atom \cite{fukai2006metal}, 
and this rule is also applicable to Fe with close-packed structures~\cite{machida2014site, machida2019hexagonal}. 
Because dhcp-FeH$_x$ contains four Fe atoms per unit (Fig.~\ref{fig:xtal}), hydrogen-induced volume expansion was converted to 8–10~\AA$^3$ per dhcp unit-cell corresponding almost 10 times the magnetic contribution. 
This exceptionally small magnetic volume contribution suggests that the directly estimated magnetostriction is smaller than previously predicted by theory.

We constructed the pressure-volume relation of dhcp-FeH$_x$ in the paramagnetic region based on previous diamond anvil cell (DAC) compression studies at 300~K~\cite{pepin2014new, hirao2004compression}
and compared the standard unit-cell volume of the ferromagnetic phase and the paramagnetic phase to evaluate $\Delta V_\text{M}$. 
The equation of state of paramagnetic dhcp-FeH$_x$ was obtained by refitting $PV$ datasets above 60~GPa, which is well above the magnetic transition pressure. 
The pressure derivative of $K_0$ ($K_{0}^{'}$) is fixed to be 4. Applying the dataset to Birch-Murnaghan equation of state, $V_0$ and $K_0$ are estimated to be 54.3(5)~\AA$^3$ and 165(9)~GPa, respectively. 
The derived unit-cell volume of paramagnetic dhcp-FeH$_x$ is significantly larger than the theoretically predicted value, 
which ranges from 50 to 51~\AA$^3$ at 300~K~\cite{pepin2014new, elsasser1998ab2}. 
It suggests that the computationally obtained volume of the paramagnetic regime is likely underestimated 
(i.e., the previously reported $\Delta V_\text{M}$ values have been overestimated) 
and the discrepancy in $V_0$ is due to a lack of low-pressure data. 
To reproduce the theoretically predicted $V_0$, the pressure derivative of bulk modulus must be set to an abnormally high value. Hirao et al. (2004)~\cite{hirao2004compression} reported $K_{0}^{'} = 8.5(29)$, 
which is twice $K_{0}^{'} = 4$ to fit the computationally predicted small $V_0$ of paramagnetic dhcp-FeH~\cite{elsasser1998ab2}. 
They also noted that if applying all of their $P$–$V$ dataset with $K_{0}^{'} = 4$, 
the resultant $V_0$ become 55.3(5)~\AA$^3$, which is near to our calculated one from the dataset. 
It implies that magnetic contribution to $V_0$ can be very small than computational results.

\begin{figure}[htbp]
    \centering
    \includegraphics[width=0.5\textwidth]{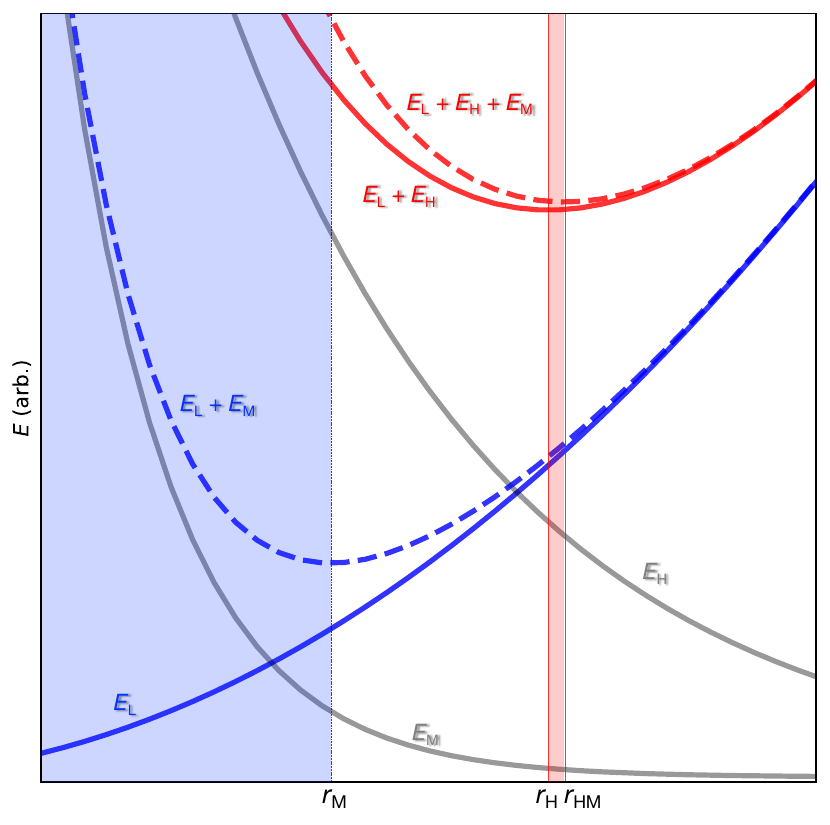}
    \caption{Schematic drawing to explain the qualitative hypothesis for the exceptionally small magnetic volume contribution of iron hydrides. 
    The blue solid line represents the relation between interatomic distance and elastic energy. 
    When the additional energy from magnetism emerges, the atomic-distance dependence of the potential-energy shift shifts to the dashed blue line. 
    On the other hand, hydrogenation also increases the interatomic distance, shifting the energy curve to the red solid line. 
    For the case of both magnetic and hydrogen effects at the same time, the energy curve shifts to the red dashed line. 
    Compared to the obtained minimum of each potential, the shift in interatomic distance for the Fe hydride (red shaded area) is smaller than that of Fe (blue shaded area).
  }
    \label{fig:schematic}
\end{figure}

One possible interpretation for this overestimation is that the volume expansion due to hydrogenation has already sufficiently increased the Fe–Fe distance, 
allowing structural stabilization even with a relatively small repulsive force from magnetostriction. 
We present a simple model illustrating the coupled effect of hydrogen-induced and magnetism-induced volume expansion. Figure~\ref{fig:schematic} shows the schematic relationship between atomic displacement ($r$) and energy ($E$) when hydrogen occupies an interstitial site. 
Hydrogen incorporation expands the Fe–Fe interatomic distance and reduces the short-range repulsion, which lowers the quantum mechanical ground state energy of the interstitial H atom ($E_\text{H}$). 
For simplicity, we treat the total energy as the sum of $E_\text{H}$ and the elastic energy of the Fe lattice ($E_\text{L}$), where EH decreases with increasing r and $E_\text{L}$ increases approximately quadratically. 
We then introduce the magnetic energy term, $E_\text{M}$, which also decreases with increasing interatomic spacing. 
The total energies of Fe and FeH including magnetic contributions are represented by $E_\text{L}$ + $E_\text{M}$ and $E_\text{H} + E_\text{L} + E_\text{M}$, giving equilibrium distances $r_\text{M}$ and $r_\text{HM}$, respectively. 
Because hydrogen pre expands the lattice, the additional expansion driven by magnetism ($r_\text{HM} - r_\text{H}$) becomes smaller than that in pure Fe ($r_\text{M}$). 
Here, we assumed that $E_\text{H}$ and $E_\text{M}$ exponentially decays with $r$ for illustration, this behavior does not rely on the specific choice of functional forms. 
Rather, it follows from general physical properties: the lattice energy is convex, and both hydrogen and magnetic energies decrease with increasing Fe–Fe distance. 
Under these broad conditions, hydrogenation naturally reduces the effective magnetostriction, and the total volume change becomes a nonlinear interplay of the two mechanisms rather than a simple sum of their individual contributions.

\clearpage

\subsection{Pressure dependence of magnetoelastic coupling of dhcp-FeH$_x$}
\begin{figure}[htbp]
    \centering
    \includegraphics[width=0.8\textwidth]{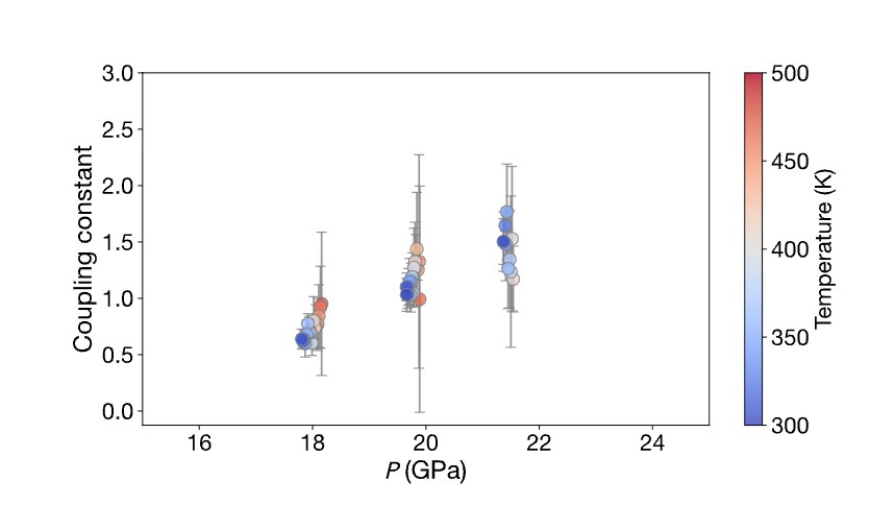}
    \caption{Pressure dependence of magnetoelastic coupling constant of dhcp-FeH$_x$. Magnetoelastic coupling constant is proportional to the spontaneous magnetostriction divided by squared magnetic moment of the ferromagnetic region. 
    The magnetic moment was estimated using experimental values at 300~K with mean-field approximation. The lowest-pressure dataset was excluded. To estimate the bulk modulus of it, the $P$–$V$–$T$ dataset above 600~K 
    --the temperature is high enough to the paramagnetic region-- was fitted to second-order Birch-Murnaghan EoS with Mie-Gr\"{u}neisen-Debye approximation. 
    Bulk modulus was calculated from the EoS of paramagnetic dhcp-FeH$_x$. Even though this assumption is relatively rough, the change in bulk modulus 
    (i.e., the change in compressibility) is too small to reproduce $T/T_\text{C}$–$\omega$.
  }
    \label{fig:P_C}
\end{figure}

Pressure dependence of magnetoelastic coupling ($C$) has not been well examined, and magnetoelastic coupling is usually regarded as constant in the Invar alloys. 
However, our observation suggests that magnetoelastic coupling effect depends on pressure. 
Figure~\ref{fig:P_C} shows the derived magnetoelastic coupling effect. 
Although the uncertainty in $C$ is very large due to the small magnetic contribution to the unit-cell volume, 
the value of $C$ at 21~GPa is approximately twice that at 18~GPa, likely showing a positive dependence on pressure. 
In this system, the temperature dependence of magnetostriction remains almost unchanged within the pressure range where experiments were conducted, while the decrease in magnetic moment due to increasing pressure is significant. 
Because the pressure and temperature dependence of bulk modulus is small to reproduce such a large pressure dependence on $C$, 
the pressure-independent magnetostriction requires enhancement of the magnetoelastic coupling constant with increasing pressure. 

This finding can explain the switching behavior from the Invar to NTE with increasing pressure. 
At the lowest pressure, the magnetoelastic contribution is insufficient to overcome normal thermal expansion, resulting in the absence of negative thermal expansion. 
As pressure increases, a distinct kink in the $T$–$V$ curve becomes sharper (Fig.~\ref{fig:PTV}B). 
Previous studies introduced the coupling constant a priori to reproduce the observed values. Our results demonstrated that pressure significantly enhances the magnetoelastic coupling in dhcp-FeH, 
leading to a crossover from nearly invariant thermal expansion to negative thermal expansion. 
This pressure dependence offers new insight into the physical significance of the coupling constant.

\clearpage

\subsection{Pressure–Temperature–Magnetization map of dhcp-FeH$_x$ revealed by experiments combined with DFT+DMFT calculations}
\begin{figure}[htbp]
    \centering
    \includegraphics[width=\textwidth]{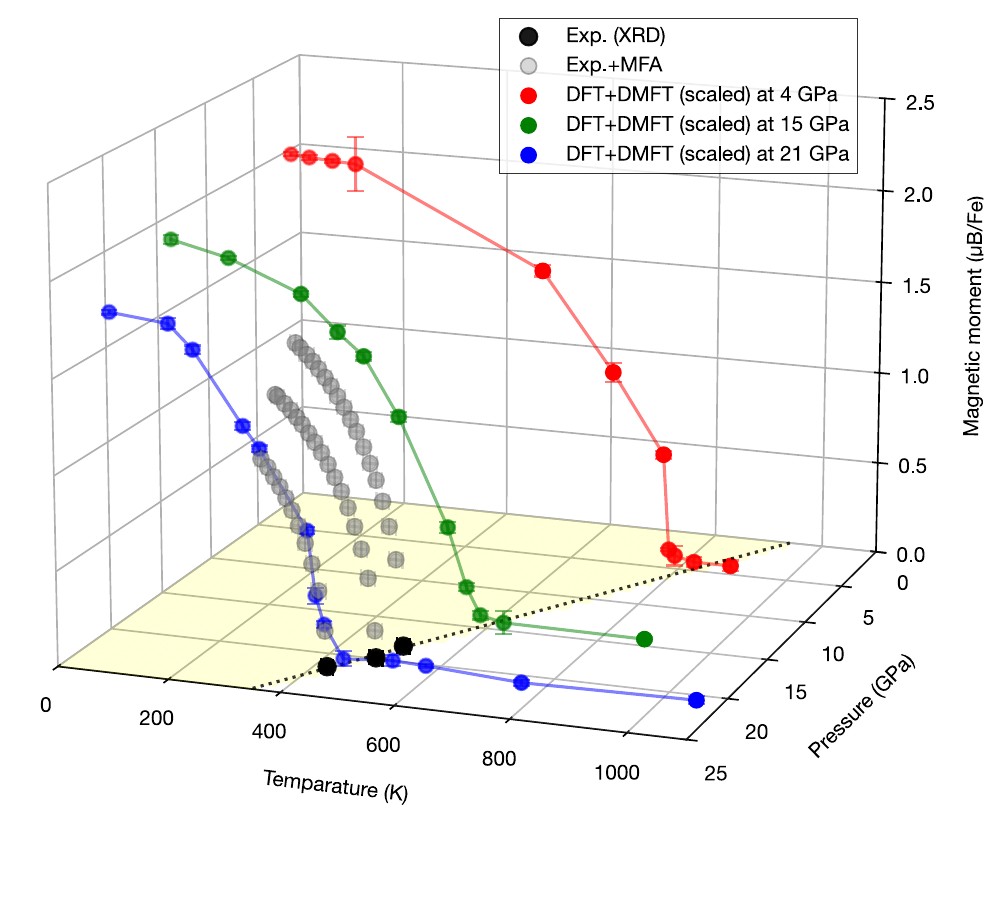}
    \caption{$P$–$T$–$M$ of dhcp-FeH$_x$ obtained from $T$–$V$ kinks and DFT+DMFT calculations. 
    Dots are magnetization, $M$, as a function of the scaled temperature at $P = 4$~(red), $15$~(teal), and $21$~(blue)~GPa (Fig.~\ref{fig:PTC_DMFT}A for unscaled data).
    Black solid circles are the experimentally determined $T_\text{C}$ and the dotted line shows the linear fit to the experimental values as a function of pressure to scale the DFT+DMFT results. 
    Grey solid circles represent the magnetic moments estimated from experimentally determined $T_\text{C}$ at high pressures (this study) and magnetic properties at 300~K (2, 27).
  }
    \label{fig:PTM}
\end{figure}

The experimentally determined $P$–$T$–$M$ relations and pressure dependence of $T_\text{C}$ are compared with DFT+DMFT calculations in Fig.~\ref{fig:PTM}.  
Using the definition of $T_\text{C}$ as the temperature where magnetization becomes zero (see arrows in Fig.~\ref{fig:PTC_DMFT}A), the calculated $P$–$T_\text{C}$ slope matches the experimentally determined ones well. 
The magnetic transition in dhcp-FeH$_x$ can be qualitatively explained by the Stoner criterion, indicating an itinerant origin of magnetism~\cite{ishimatsu2012hydrogen}. 
Within the itinerant electron model of ferromagnetism, the pressure dependence of $T_\text{C}$ is described as a sum of a positive contribution associated with strong ferromagnetism and a negative contribution characteristic of weak itinerant ferromagnetism~\cite{edwards1968magnetic, kiss2013pressure}. 
The observed negative pressure dependence of $T_\text{C}$ suggests that the weak itinerant term associated with spin fluctuations plays a significant role. This interpretation is consistent with computational results, which indicate that dhcp‑FeH exhibits a dual itinerant–local character.

At the highest (29~GPa) and lowest (4~GPa) pressures at which DMFT calculations were performed, the linearity of the relationship between pressure and $T_\text{C}$ is expected to collapse. At 4 GPa, the magnetic moment is almost saturated, revealing $M_\text{S}$~2.06~µB/Fe-atom at $T$ = 116~K. 
The obtained magnetic moment of dhcp-FeH$_x$ is comparable to previous DFT calculations and ND experiments\cite{saitoh2020crystal, tsumuraya2012first}. 
Our DFT+DMFT calculations revealed the temperature dependence of the magnetization deviated from the typical Ginzburg–Landau behavior at 29~GPa, 
indicating a strong enhancement of spin fluctuations. 
This observation is well in agreement with XMCD and M\"{o}ssbauer measurements~\cite{ishimatsu2012hydrogen, mao2004nuclear}.

Complementary research of high-$PT$ XRD and DFT+DMFT calculations reproduced the observed pressure dependence of the magnetic properties of dhcp-FeH$_x$. 
While DFT+DMFT calculations have recently been increasingly applied to ferromagnetic metals such as bcc-Fe~\cite{han2018phonon, belozerov2014coulomb, belozerov2017momentum} and hcp-Fe~\cite{pourovskii2020electronic}, 
this study represents the first application to dhcp-FeH$_x$ in a wide $PT$ range. Our result showed that dhcp-FeH$_x$ serves as an ideal benchmark system for validating computational methods owing to its structural stability over a wide density range and its well-defined ferromagnetic–paramagnetic transition.

\clearpage
\section*{Data and materials availability} 
The crystal structure of dhcp-FeH$_x$ in Fig.~\ref{fig:xtal} is drawn with VESTA programme~\cite{momma2011vesta}.
All data that support the findings of this study are included within the article and the obtained unit-cell volume ($P, T$) are listed on https://zenodo.org/records/15779440.

\section*{Acknowledgments}
In situ X-ray observation experiments of dhcp-FeH$_x$ were performed at BL04B1, SPring-8 and NE7A, PF-AR, KEK 
(project numbers: 2023B0314, 2024A0314 to Y.M., 2023G668 to H.K.). 
High-pressure cell for metal hydrides was developed at BL04B1, SPring-8 
(project numbers: 2022A1319, 2022A2067, 2022B2118, 2023A1250 to S.K.). 
I.P. and D.Y.K. thank Ji Hoon Shim, Junwon Kim, and Hong Chul Choi (POSTECH) for fruitful discussion.

\section*{Competing interests}
All other authors declare they have no competing interests.

\section*{Author contributions}
Y.M. conceived and designed the research. 
Y.M., M.T. and H.K. performed X-ray diffraction experiments.
Z.W. and D.Y.K. conducted preliminary DFT calculations, and I.P. and D.Y.K. performed DFT+DMFT calculations. Data curation and visualization was done by Y.M. 
All authors reviewed and approved the manuscript before submission.


\clearpage
\bibliographystyle{vancouver} 
\bibliography{references} 





\clearpage
\setcounter{figure}{0}
\renewcommand{\thefigure}{S\arabic{figure}}

\section*{}

\begin{figure}[]
    \centering
    \includegraphics[width=\textwidth]{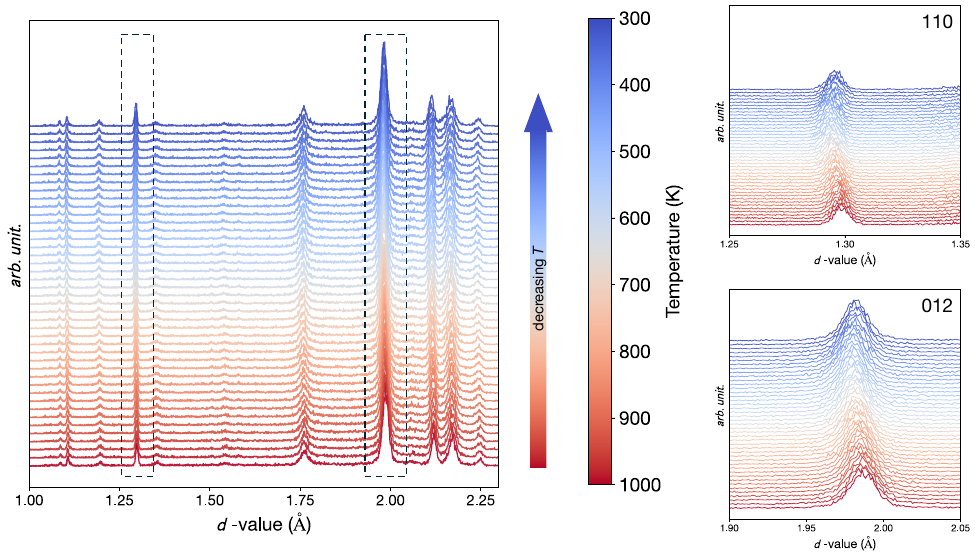}
    \caption{The stacked XRD profiles during the cooling process at 21~GPa. 
    The negative thermal expansion was most pronounced at this pressure. 
    The two peaks were magnified for the clarity: 012, which corresponds to the strongest X-ray diffraction peaks of dhcp iron hydride, 
    and 110, which exhibits a very sharp peak.}\label{figS:Index}
\end{figure}
\clearpage

\begin{figure}[]
    \centering
    \includegraphics[width=\textwidth]{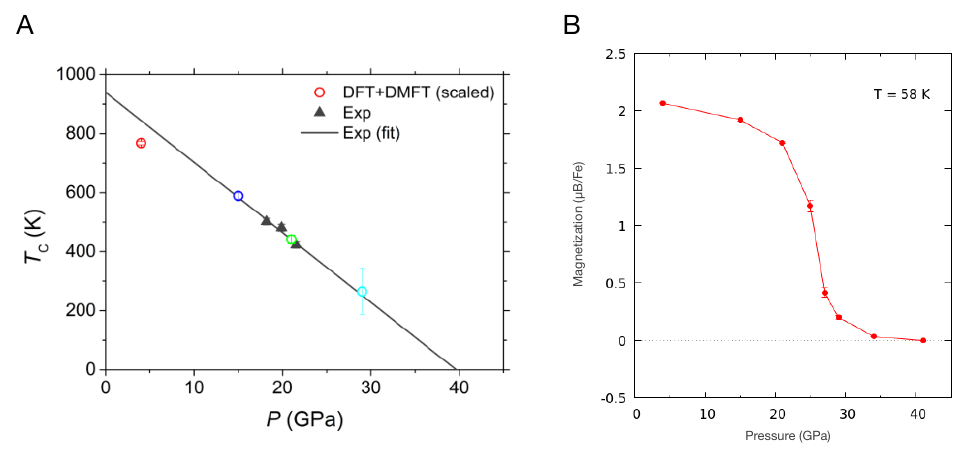}
    \caption{Pressure dependence of $T_\text{C}$ and magnetization extended to $P$ = 29~GPa. 
    (A) $P$–$T_\text{C}$ plot up to 29~GPa. 
    (B) The temperature condition is $T$ = 58~K except for $P$ = 4~GPa. 
    For $P$ = 4~GPa, $T$ = 116~K was used, at which the magnetization was saturated. 
    The vertical error bar indicates the standard deviation of magnetization measured for the last 10 converged Monte Carlo calculations.}
    \label{figS:29GPa}
\end{figure}
\clearpage

\begin{figure}[]
    \centering
    \includegraphics[width=\textwidth]{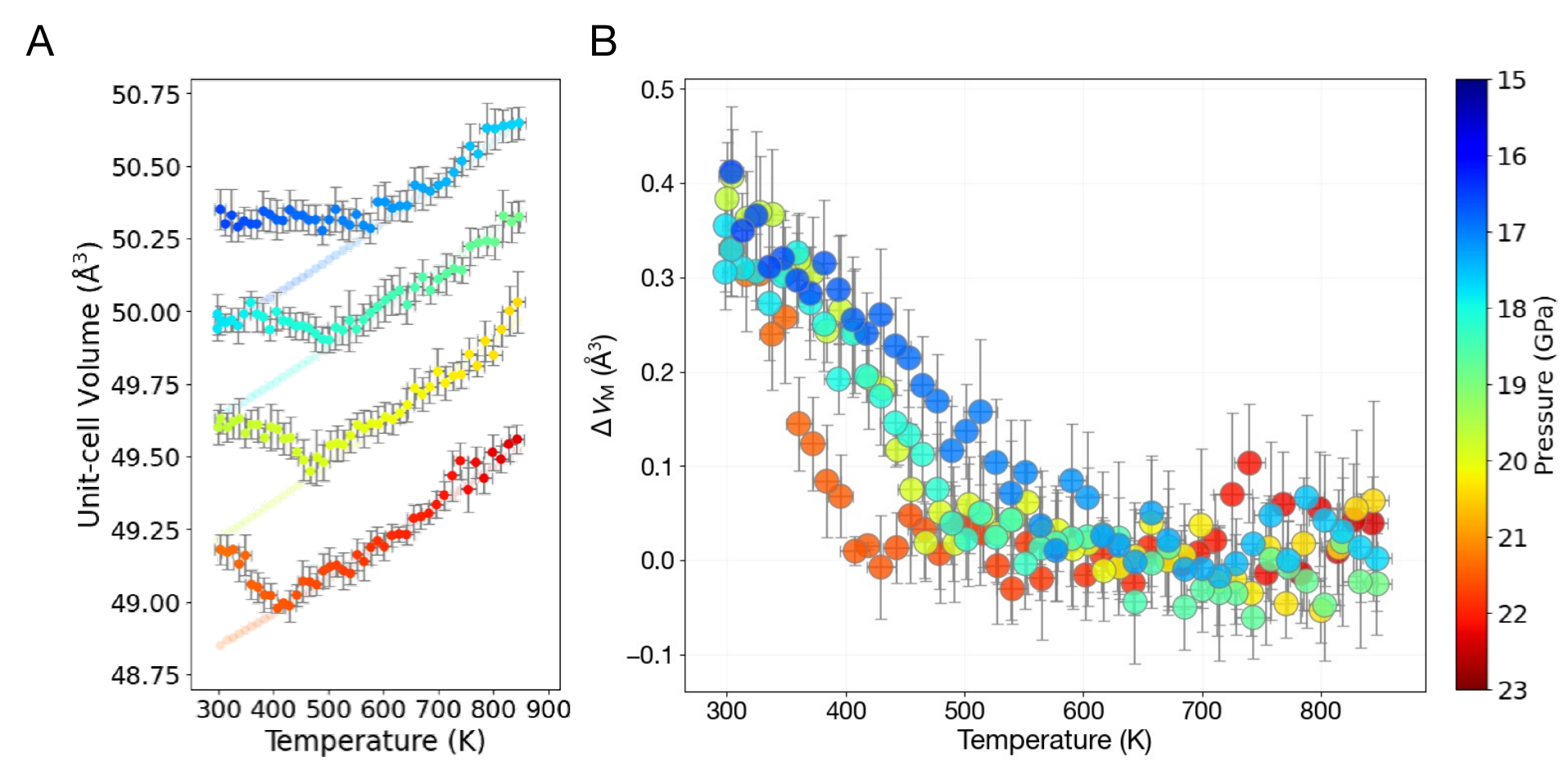}
    \caption{
    (A) $T$–$V$ relation of dhcp-FeH$_x$ at each $PT$ points. 
    In addition to the measured $T$-$V$ relationship (Fig.~\ref{fig:PTV}B),
    the volume at each $PT$ point is also plotted using the EoS of the ferromagnetic constructed above 600~K.
    (B) The difference between the observed unit-cell volume and the calculated paramagnetic volume at each $PT$ point corresponds to the magnetic contribution to the volume ($\Delta V_\text{M}$).
    A graph with the temperature on the horizontal axis normalized by $T_\text{C}$ is shown in Fig.~\ref{fig:T_Vmag}.}
\label{figS:Vmag_diff}
\end{figure}
\clearpage

\begin{figure}[]
    \centering
    \includegraphics[width=\textwidth]{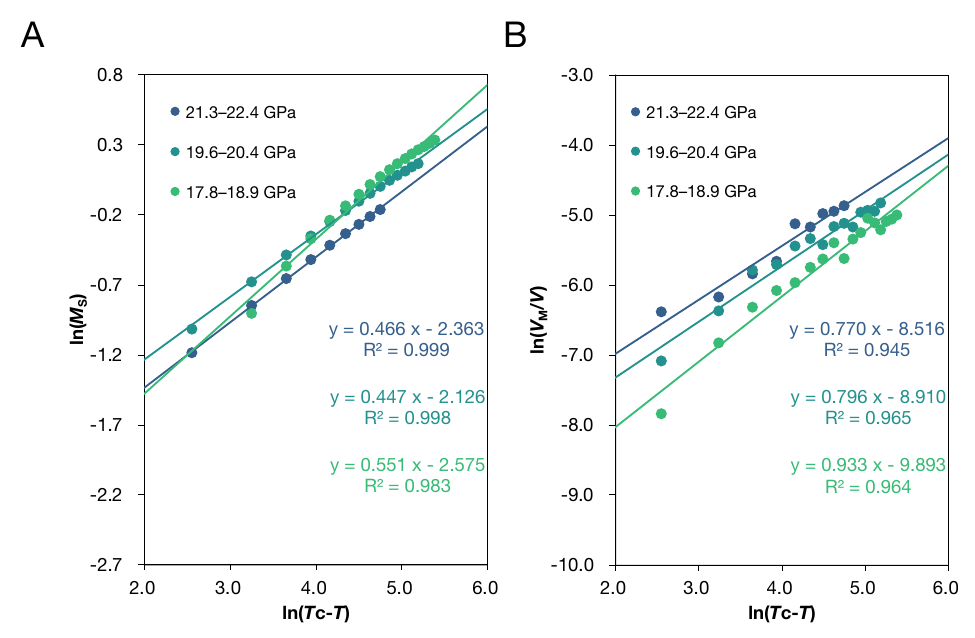}
    \caption{The critical exponent of $T$–$\Delta V_M$ and $T$–$\omega$. 
    The critical exponent estimated from the experimentally obtained $T_\text{C}$ and mean-field approximation 
    and the experimentally obtained $\Delta V_M$  and the relation between spontaneous volume magnetostriction ($\omega$) and spontaneous magnetization: 
    (A) The relation between $M_\text{S}$ and $T/T_\text{C}$ of each path. $M_\text{S}$  at room temperature was estimated using the pressure dependence of magnetic moment of Fe in dhcp-FeH$_x$ clarified by the previous studies~\cite{ishimatsu2012hydrogen, saitoh2020crystal}
    and mean-field approximation with $S = 1/2$. The slope is corresponding to the critical exponent, $\beta$. Note that $M_\text{S}$–$T/T_\text{C}$ relation is described to $T/T_\text{C}\sim$0.6, 
    which exceeds ‘the vicinity’, and the modeled slope is slightly deviated from 0.5. 
    (B) The relation between $\Delta V_M$  and $T/T_\text{C}$ of each path. 
    The spontaneous volume magnetostriction ($\Delta V_M$) is proportional to the square of the spontaneous magnetization ($M_\text{S}$ ). 
    The slope of these plots corresponds to 2$\beta$.}
\label{figS:critical}
\end{figure}
\clearpage

\setcounter{table}{0}
\renewcommand{\tablename}{Table}
\renewcommand{\thetable}{S\arabic{table}}

\begin{table}
\caption{Crystal structure of dhcp-Fe used in DMFT calculation. The unit-cell volume ($V$) is directly obtained from the reference \cite{hirao2004compression}, and the ($c/a$) ratio is 3.270.}
\vspace{5 mm}
\centering
\begin{tabular}{c c c}
\hline
Pressure (GPa) & Volume(\AA$^3$) & Lattice constant, $a$\\
\hline
4       & 53.78  & 2.668  \\
15      & 50.35  & 2.610  \\
21      & 48.86  & 2.584  \\
29      & 47.24  & 2.555  \\
\hline
\end{tabular}
\label{tab:dmft-structure}
\end{table}

\end{document}